# Electro-Optical Switching in a Nonlinear Metasurface

Matthew Feinstein[1,2], Alexander Andronikides[1] and Euclides Almeida[1,2,*]

[1]*Department of Physics, Queens College, City University of New York, Flushing, NY 11367, United States of America*

[2]*The Graduate Center of the City University of New York, New York, NY 10016, United States of America*

*Author's email address: euclides.almeida@qc.cuny.edu

**Abstract**

Nonlinear metasurfaces are multifunctional photonic elements that generate and control light, enabling multiple proof-of-principle applications, such as in nonlinear holography, beam shaping, and nanoscale sources of entangled photon pairs. Active tuning of nonlinear metasurfaces will considerably expand their prospects for integrated photonics. Here, we demonstrate a highly-tunable metasurface device that combines electrical and optical switching in a broadband frequency conversion process. Our device employs electrically tunable hybrid graphene-gold plasmons to convert mid-infrared pulses into visible light by optical gating. A strong electrical modulation of the generated visible light is attained for a broad range of mid-infrared input frequencies, and this modulation is up to 3.5x stronger for the metasurface than for plain graphene. All-optical switching measurements indicate that the device can be optically switched at Terahertz rates. Our results may lead to new applications of metasurfaces such as ultrafast signal processing, nonlinear sensing, and optical transduction.

**Introduction**

Nonlinear metasurfaces have emerged as versatile multifunctional photonic elements with potential applications as compact light sources [1-7]. They also serve as a rich playground for fundamental studies of nonlinear light-matter interaction [3, 8, 9]. Unlike most conventional linear metasurfaces, they allow for simultaneous generation and manipulation of coherent light [10]. Parameters such as polarization, phase and amplitude of the frequency-converted light can be controlled locally with a high degree of precision [3, 4, 6, 11, 12]. Proof-of-principle applications of nonlinear metasurfaces include nonlinear beam shaping [13] and holography [14, 15], and generation of quantum correlated photon pairs [16, 17]. Since nonlinear optics can connect disparate electromagnetic spectrum frequencies, nonlinear metasurfaces can cover a wide range of frequencies, from Terahertz [18] down to deep [19, 20] and vacuum [21] ultraviolet.

Active tuning of nonlinear metasurfaces may greatly expand their viability for integration with photonic devices [22, 23]. However, conventional metasurfaces are composed of metallic or dielectric nanostructures which both exhibit limited dynamic tuning. Noble metals have a large density of electrons in the conduction band. Therefore, external voltages barely change the electronic density and consequently the plasmonic resonances. In dielectric nanoresonators, external voltages will primarily induce bound charges, which will not cause significant changes in their photonic resonances that depend on the refractive index.

Large tuning of the optical nonlinear response can be attained in atomically thin materials such as graphene [24-27] and transition metal dichalcogenides [28], which have remarkably strong nonlinearity for their thinness [29-32]. In graphene, the mechanism for nonlinearity tuning occurs through the injection of charge carriers, which significantly alters its Fermi level, affecting both the interband and intraband transitions [24, 26]. Previous studies have shown that electrical gating can greatly modify graphene's third-order nonlinearity for different frequency mixing processes [24, 26]. Moreover, in nanostructured graphene, plasmons are widely tunable across the terahertz and the mid-infrared regimes [33], and this property can be used to enhance and actively control nonlinear frequency conversion [34, 35].

All-optical switching in metasurfaces enables much faster tunability than electric gating since the temporal response is not limited by the RC time constant of electrical circuits. For nonlinear metasurfaces, all-optical control may lead to ultra-compact optical limiters, saturable absorbers, ultrafast switches, and optical transistors, with applications ranging from ultrafast lasers to optical information processing [22].

Here, we demonstrate a multifunctional nonlinear metasurface that combines electrical and optical switching in a single device. Our approach uses electrically tunable hybrid graphene-gold plasmons to convert mid-infrared pulses into visible photons. Broadband frequency conversion is accomplished by an optically gated scheme, where a near-infrared beam incident on the metasurface interacts with a tunable mid-infrared pulse by four-wave mixing. By electrically tuning the near-field of the metasurface, the generated nonlinear signal can be modulated up to 88% over a wide band of mid-infrared frequencies, and modulation is 3.5x times larger than that of bare graphene. Moreover, time-delayed measurements suggest an ultrafast optical response – nearly 200 fs – limited by the lifetime of the hybrid gold-graphene plasmons. Our work represents a significant step for developing electro-optical metasurfaces that convert light from contrasting regions of the electromagnetic spectrum, with potential applications in optical information processing and quantum transduction.

**Results**

Figure 1(a) illustrates the electro-optical tunable metadevice and the optical scheme to convert mid-infrared pulses into visible light. The metasurface consists of an array of gold nanorods on top of graphene arranged in a back-gated field-effect transistor configuration. In this metasurface, mid-infrared radiation couples to hybrid graphene-gold plasmons [36], which can be switched on and off by the back-gate voltage. A pulsed mid-infrared radiation is converted into visible light using a gating femtosecond (150 fs) near-infrared pulse centered at λ = 800 nm. Through interaction with the third-order nonlinearity of graphene, a four-wave mixing signal is generated at the frequency $\omega_{FWM} = 2\omega_{MIR} + \omega_{800nm}$, where $\omega_{MIR}$ is the frequency of the mid-infrared beam. The energy diagram of this energy-conserving interaction is shown in Fig. 1(b).

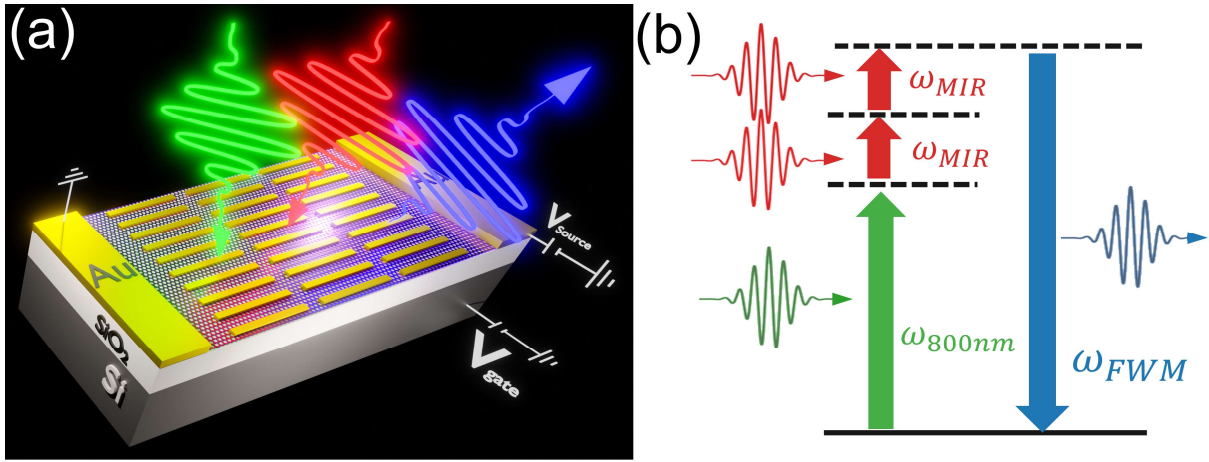

**Figure 1. Electro-optical tunable nonlinear metasurface**. (a) Illustration of the metasurface device and the mixing process to generate visible light from infrared pulses. The femtosecond 800 nm gate (green) and tunable mid-infrared (red) pulses are overlapped in space and time to achieve a strong visible FWM signal (blue), which is detected in reflection. The device consists of nanoscale gold rods on a graphene surface. The graphene layer is charged through application of a gate voltage to the silicon which is spaced from graphene by silica, forming a capacitor. (b) A representative four-wave mixing energy level diagram. One photon from the gate pulse combines with two mid-infrared photons on the metasurface to generate the visible photon with frequency $\omega_{FWM} = 2\omega_{MIR} + \omega_{800nm}$.

The design of the metasurface is shown in Figure 2(a), along with its geometric parameters. Adjusting the gate voltage injects carriers onto graphene, fine-tuning its Fermi level. Fig. 2(b) shows a SEM image of the fabricated gold rods on graphene, which are nominally 1210 nm long and 70 nm thin. The device exhibits graphene-like plasmons at 7.5 and 11.5 μm, which can be observed in the transmission modulation $(1 - T/T_{CNP})$ spectra in Fig. 2(c) as the Fermi level ($E_F$) of graphene increases. Fig. 2(d) shows charge density maps at the graphene-like plasmon resonance at 7.5 μm. The right panel shows doped graphene (0.35 eV) with charge oscillations concentrated on the graphene between the side edges of the gold rods, corresponding to electric fields concentrated on sections of graphene. This graphene-like plasmon can be disabled by tuning the graphene near the charge neutral point (CNP, $E_F = 0\ eV$) of graphene (Fig. 2d, left panel).

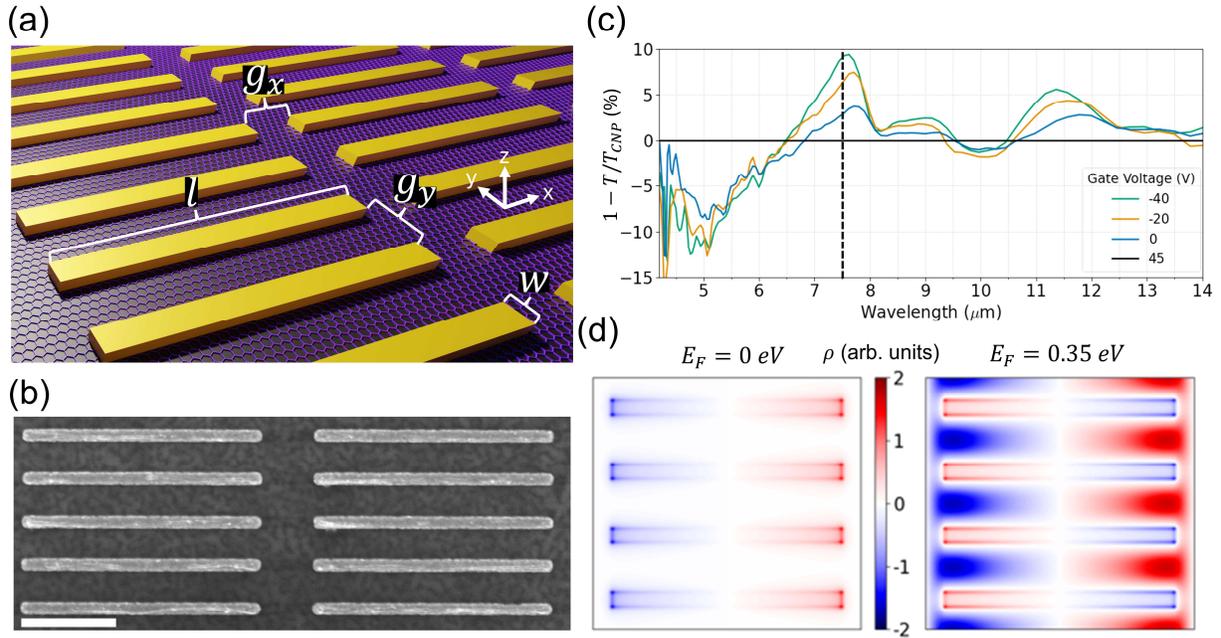

**Figure 2. Device Design and Linear Measurements.** (a) Schematic of metasurface nanostructure. Gold rods with dimensions of l = 1230 nm and w = 70 nm) are arrayed on the graphene with gaps of $g_x$ = 260 nm along the long axis and $g_y$ = 150 nm along the short axis. (b) SEM image of the gold rods (brighter) on graphene (darker). Scale bar is 500 nm. (c) Plot of measured spectra of transmission modulation, defined as $1 - T/T_{CNP}$ where $T_{CNP}$ is the transmission at the charge neutral point. Spectra for different Fermi energies are plotted, and a clear increase in transmission modulation is seen around 7.5 and 11.5 µm, corresponding to graphene-like plasmon resonances. The black dashed vertical line corresponds to the wavelength of the charge maps in (d). (d) Calculated charge density oscillation map at 7.5 µm excitation at the CNP (left panel) and at higher doping (0.35 eV) (right panel). At higher doping, strong charge oscillations are concentrated on the graphene between the tips of the gold rods, indicating a graphene-like resonance.

Figure 3 summarizes the results for the mid-infrared to visible frequency conversion process in the nonlinear metadevice. We first characterize the polarization dependence of the four-wave mixing signal. The polarization configuration of both input beams is shown in Fig. 3(a). The polarization of the mid-infrared beam is fixed parallel to the long axis of the gold rods. In contrast, the polarization of the 800 nm beam is allowed to rotate. Fig. 3(b) shows the polarization dependence of the FWM signal of the metasurface and bare graphene, with the mid-infrared beam centered at 7.5 µm.

Interestingly, in graphene, the signal reaches its maximum for cross-polarized excitation, being 3.5x stronger than that for co-polarized excitation. Moreover, at cross-polarized excitation, we found a negligible background signal (see Supplementary Information) from the silica-silicon substrate. This is an important experimental observation for nonlinear generation in graphene by wave-mixing: measuring the nonlinear signal under cross-polarization emphasizes the response of graphene while suppressing the background signal of the substrate. However, the situation is reversed in the metasurface, and the signal measured at the zero-order of diffraction is stronger for co-polarized excitation, possibly due to constructive interference between the signals from the background and graphene. Our measurements will be performed at cross-polarization to suppress the coherent background.

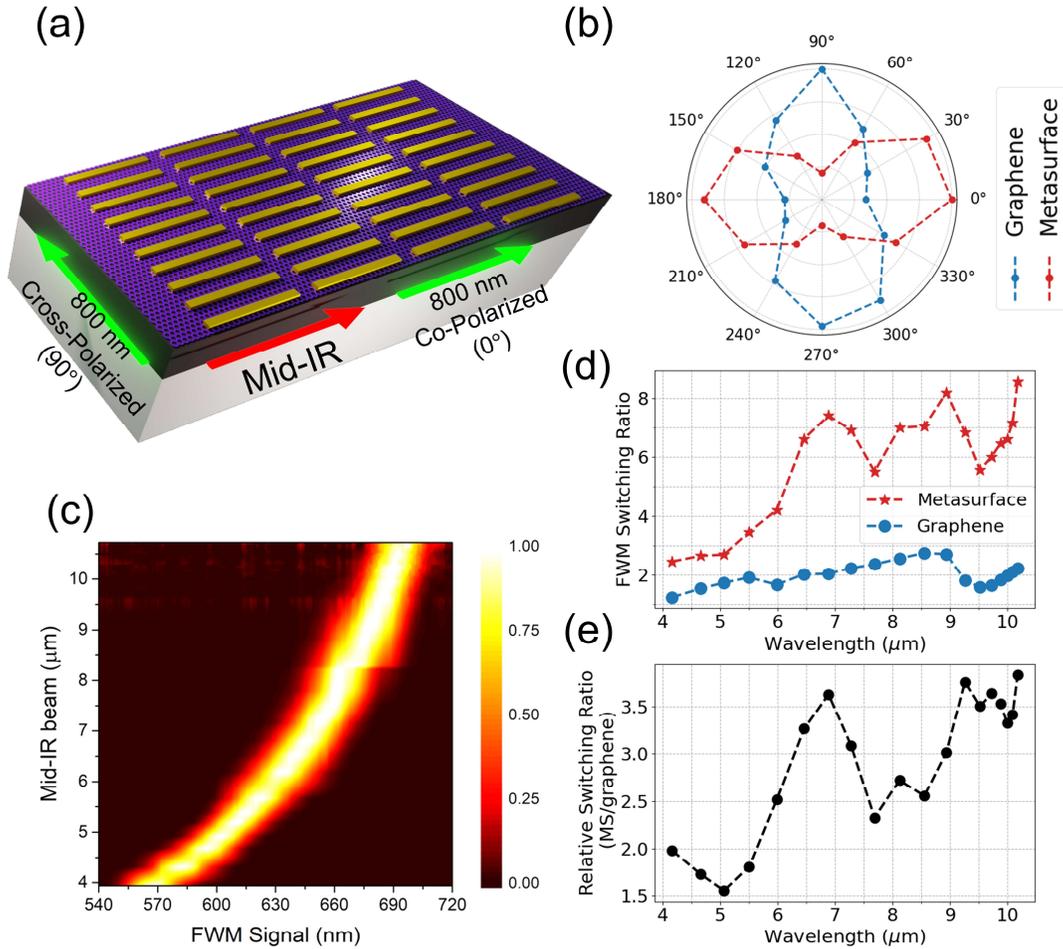

**Figure 3. Electrically Tunable Four-Wave Mixing Experiment.** (a) Schematic showing the polarization of the input beams. The mid-IR beam is polarized along the long axis of the gold rods, and the 800 nm beam polarization can be rotated. (b) The nonlinear signal plotted as a function of the 800 nm polarization angle for both unstructured graphene and the metasurface, normalized to fit on the same scale. 0° is parallel with the mid-IR beam. Unstructured graphene has the strongest signal under cross-polarization (90°), while the metasurface has the strongest signal under co-polarization (0°). (c) FWM spectra for different mid-infrared beams, normalized to each spectrum's maximum. (d) FWM switching ratio over a range of mid-infrared inputs for the unstructured graphene and the metasurface. The switching ratio is the ratio of the FWM signal at high doping to the signal at low doping. Data is for cross-polarized input beams. (e) Relative switching ratio, obtained by normalizing the switching ratio of the metasurface by graphene. The metasurface appears to have features roughly corresponding to the tunable linear transmission in Fig. 1(c).

Figure 3(c) shows the spectra of the measured FWM signal for different mid-infrared input frequencies. The FWM signal falls in the visible spectrum, between 570 and 710 nm, for the range of mid-infrared input frequencies between 4 and 10.5 μm accessible to our experimental system. This FWM signal is sensitive to voltage tuning of the mid-infrared near-fields, since the nonlinear polarization $P_{NL}$ scales as $P_{NL} \propto$

$E^2(\omega_{MIR}) \times E(\omega_{800nm})$. We expect an enhanced signal when graphene is charged, as the graphene-like plasmon concentrates the mid-infrared electric field on graphene, resulting in greater field confinement. Ideally, the fields must be confined to graphene rather than the gold, since graphene is the stronger nonlinear material [35]. The FWM signal for an identical gold rod array with no graphene was below the detection limit of our apparatus. In addition, we may expect a change in the intrinsic graphene nonlinearity as the Fermi level changes. Therefore, switching the graphene Fermi level between an "on" and "off" state will cause a substantial modulation of the nonlinear signal.

Figure 3(d) plots the "switching ratio" of the four-wave mixing signal, which is defined as the ratio of the nonlinear signal at high doping (0.45 eV) to the signal at low doping (0.15 eV). The switching ratio data is collected and plotted over the range of mid-infrared inputs for both the metasurface (the rods on graphene) and bare graphene. Bare graphene exhibits a switching ratio between 1x and 3x, indicating that the intrinsic graphene nonlinearity modestly increases with the doping at some mid-infrared frequencies. The metasurface has a higher switching ratio than bare graphene over the entire range, reaching a maximum switching ratio of approximately 8x. Notably, the metasurface switching ratio increases sharply as the wavelength is increased toward the graphene-like plasmon resonance centered at 7.5 μm. Fig. 3(e) highlights the plasmonic enhancement by normalizing the switching ratio of the metasurface by that of the plain graphene, where a sharper peak near 7 μm can be seen. Since in our experimental setup, the mid-infrared input beam is relatively broad (FWHM = 1.3 μm at $\lambda_{MIR} = 8$ μm), the wavelength dependence in Figs. 3(d) and (e) may appear more uniform than expected if narrower pulses were used.

To characterize the optical switching dynamics of the device, we conducted a cross-correlation experiment, as illustrated in Fig. 4(a). The delay between the mid-infrared beam and the 800 nm gate pulse is tuned by a motorized delay line placed on the optical path of the gate pulse. Fig. 4(b) shows the results of the cross-correlation experiment for the metasurface at two wavelengths: 4.1 μm, corresponding to the peak of the gold rod plasmonic resonance, and 7.3 μm, corresponding to the graphene plasmonic resonance. The measurements were taken for two voltage states: 40 V, near the charge neutral point, and -10 V. The data was taken for cross-polarization of the two beams to minimize the substrate background.

We extract the FWHM of each curve by fitting the data to a Gaussian, shown on each plot in Fig. 4(b). At 4.1 μm, the gate voltage causes a small shift in the correlation profile (16 fs) and a slight increase in the FWHM (10 fs). At 7.3 μm, the temporal dynamics of the metasurface has greater sensitivity to the voltage, showing a tunable FWHM (between 185 and 212 fs) and a 32 fs shift in the maximum FWM signal. For comparison, the FWHM of the convolution of the input beams is estimated to be 170 fs. This was obtained using FWHM values measured for the 800 nm beam (140 fs) and for the mid-infrared at 8000 nm (68 fs) and assuming a convolution of Gaussian beams, see Supplementary Information.

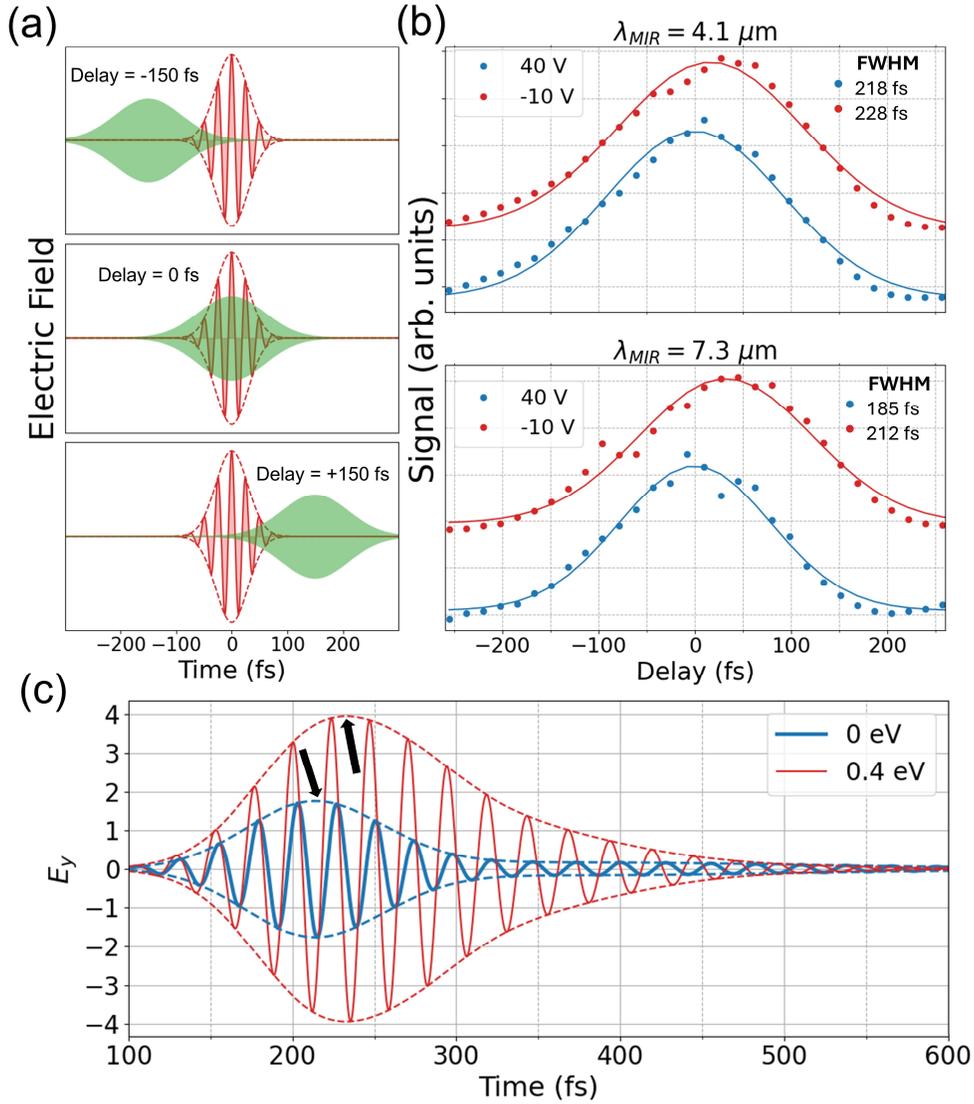

**Figure 4. Cross-Correlation of Input Beams.** (a) Illustration of the cross-correlation experiment. The mid-infrared pulse (red, FWHM = 68 fs) arrives at the sample at time = 0, while the 800 nm pulse (green, FWHM = 140 fs) can arrive with a variable delay. As the delay shifts away from zero, the overlap of the beams decreases, and the FWM signal is reduced. (b) Results of cross-correlation experiments for the metasurface for input beams of 4000 nm (top) and 7500 nm (bottom). Data (dots) is shown for both doping states of the graphene, and are fitted with Gaussian curves (solid line). The data is taken with cross-polarization of the beams. The peak of the 40 V curve is taken as delay = 0 fs. The curve magnitudes are made to the same scale to emphasize the FWHM change and peak shift, and are arbitrarily offset along the signal axis for visibility. (c) Calculated y-component of electric field (solid lines) with envelope (dashed lines) on graphene over time at the CNP (blue) and when doped (red) for an excitation centered around 7.3 μm. The peaks of each curve are denoted by arrows, and the peak at higher doping is delayed by ~20 fs. In addition, the optical fields persist for much longer at higher doping.

The larger FWHM measured at higher doping is understood to be due to the plasmonic resonances of the metasurface. At 40 V, near the charge neutrality point, the gold nanorod resonance dominates for both 4.1 μm and 7.3 μm excitations. At -10 V, however, the mid-infrared pulse couples to graphene plasmons at 7.3 µm, changing the dynamics of the FWM signal generation. As the graphene plasmons resonate, we expect the signal to remain for longer delays between the beams since the sample retains the plasmonic near-fields even after one beam has passed, resulting in a larger FWHM. This is illustrated in Figure 4(c), where the electric field on graphene at 7.3 µm is calculated for $E_F = 0$ eV and $E_F = 0.4$ eV, and the near-fields are seen to persist on the graphene for much longer at higher doping. In addition, there is a small shift (~20 fs) in the peak magnitude of the electric field oscillations. The ~30 fs increase in the delay at higher doping observed in the experiment is related to both the shift in the peak observed and the increased persistence of the graphene plasmon observed in Fig. 4(c) and is explained in the Supplementary Information by considering the convolution of the 800 nm pulse with the electric field dynamics shown in Fig. 4(c). At 4.1 µm, however, graphene plasmons are damped by optical phonons [37], and its influence on the cross-correlation profile is minimized. In this case, the gold rod resonance dominates the cross-correlation profile, which is only slightly affected by the gate tuning.

Consideration of the delay curves in Fig. 4(b) implies that the switching ratio, defined above as the ratio of the FWM signal at high to low doping, will depend on the delay between the beams. The measurements in Fig. 3 were performed with the delay line adjusted for maximum signal at the higher doping condition. By choosing a delay condition further from the maximum signal where the lower doped signal is significantly decayed but the higher doped signal is on the same scale as its peak value, the switching ratio can be enhanced at the cost of overall signal strength.

*Theoretical modeling of the FWM signal*

Our metasurface device takes advantage of the field confinement of tunable graphene plasmons to enhance the modulation by a factor of up to 3.5x that of bare graphene. Looking at the simulated linear near-fields can provide an estimate of the tunability of the graphene FWM response [38]. The nonlinear polarization $P_i^{NL}(\omega_{FWM})$ can be expressed as a product of the electric near-fields at the metasurface

$$P_i^{NL}(\omega_{FWM}) \propto \sum_{j,k,l} \chi_{ijkl}(\omega_{FWM}; \omega_{MIR}, \omega_{MIR}, \omega_{800}) E_j(\omega_{MIR}) E_k(\omega_{MIR}) E_l(\omega_{800}) \quad (1)$$

where the subscripts (i,j,k,l) refer to polarizations *x* or *y*, $\chi$ is the nonlinear susceptibility tensor, and $E$ are the electric fields at the relevant frequencies. For graphene, $\chi$ is simplified, as the symmetry of graphene reduces the in-plane components to only 8 non-zero terms [29], consisting of $\chi_{xxxx}^{(3)}$, $\chi_{xyxy}^{(3)}$ and $\chi_{xxyy}^{(3)}$, along with their interchanges ($x \leftrightarrow y$) and cycles.

In the case of measurements on unstructured graphene, the mid-infrared field components are only along the incident polarization direction (referred to as *x*). This means that when we measure the FWM signal of bare graphene, we select only $\chi_{xxxx} = \chi_{yyyy}$ when the gate pulse ($\lambda$ = 800 nm) is co-polarized with the mid-infrared beam, and $\chi_{yxxy} = \chi_{xyyx}$ when the beams are cross-polarized. Since the ratio between the co- and cross-polarized signals was measured to be around 3.2 (at 0 V), the ratio between the susceptibility components is $|\frac{\chi_{xyyx}}{\chi_{xxxx}}| = \sqrt{3.2}$.

When measuring the metasurface, the mid-infrared beam is always polarized along one direction (referred to as $x$), and the metasurface confines the mid-infrared field, producing components along both directions, which makes all terms potentially relevant. In practice, the susceptibility terms $\chi^{(3)}_{xyxy} = \chi^{(3)}_{yxyx}$ and $\chi^{(3)}_{xxyy} = \chi^{(3)}_{yyxx}$ have little contribution to the cross-polarized FWM signal, as is demonstrated with calculated near-field maps in the Supplementary Information. Therefore, the nonlinear polarization can be considered as a function of terms involving $E_x^2(\omega_{MIR})$ and $E_y^2(\omega_{MIR})$. The near-field component $E_y(\omega_{MIR})$ is expected to increase substantially with increasing the Fermi level as the graphene plasmon kicks in [36]. The switching ratio of the FWM signal can be estimated by considering the far-field by the overlap integral

$$\hat{x}_i \cdot \vec{E}_{far}(\omega_{FWM}) \propto \int dV \, \vec{P}_{NL} \cdot \vec{E}_{near}(\omega_{FWM}) \tag{2}$$

where $\vec{E}_{far}(\omega_{FWM})$ is the electric field of the FWM signal in the far-field, and $\vec{P}_{NL}$ is the nonlinear polarization in Eq. (1). $\vec{E}_{near}(\omega_{FWM})$ is the near-field at $\omega_{FWM}$ on the metasurface excited by a plane wave with polarization vector $\hat{x}_i$ launched from the detector towards the metasurface. $V$ is the volume of the nonlinear material, which we confine to the surface of graphene in our analysis to gain insight into the tunable signal generated by our metasurface. The intensity of the signal is proportional to the square modulus of this overlap integral. This analysis is based on the Lorentz reciprocity theorem, and the Supplementary Information of Ref. 38 provides a detailed derivation and explanation.

We calculate these values using the simulated mid-infrared and optical fields and present the results in Fig. 5 for the case of cross-polarized light. Figure 5(a) shows the calculated overlap fields $\vec{P}_{NL} \cdot \vec{E}_{near}$ at both low and high doping under 7 $\mu$m mid-infrared excitation. The ratio of FWM signal intensities at high and low doping across the range of mid-infrared input frequencies is shown in Fig. 5(b). The calculations support the intuition that the switching ratio is strongest in the presence of the graphene plasmons, which can be seen to contribute greatly to the overlap fields in Fig. 5(a). Fig. 5(b) can be compared to the experimental data in Fig. 3(e). While the quantitative values differ substantially, the theory and experimental plots share some features, such as a prominent peak around 7 $\mu$m, a bump around 8 $\mu$m, and a wide feature between 9 and 10.5 $\mu$m. This calculation does not account for contributions of fields in the z-direction, and only adds up fields in the graphene plane. In addition, we assume that the susceptibility components are in phase with each other, though in reality they may interfere and the phase may even depend on the graphene doping.

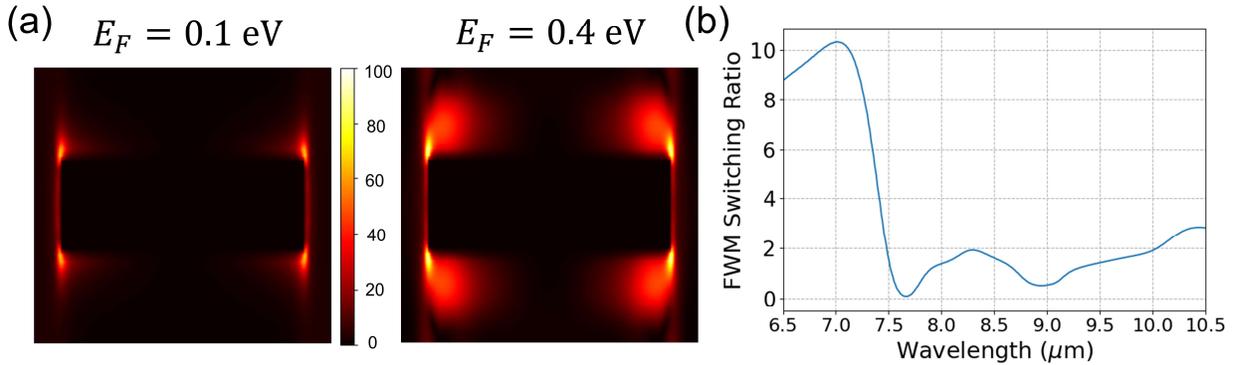

**Figure 5. Calculated Nonlinear Polarization and Far-field Signal.** (a) Calculated overlap fields $\vec{P}_{NL} \cdot \vec{E}_{near}$ of the FWM process with 7 µm mid-infrared input and cross-polarized conditions. Each column represents a different Fermi level of graphene, corresponding to the two Fermi levels used in the experiment. b) Calculated FWM signal switching ratio between the two Fermi level, computed from the overlap integral, as a function of the mid-infrared input wavelength.

**Discussion**

We have presented a tunable, broadband electro-optical nonlinear metadevice that converts mid-infrared radiation into visible light. An optically-gated scheme converts mid- and long-wave infrared pulses into visible photons within the green-to-red range (560-700 nm). Electric switching is attained by tuning the near-field of hybrid graphene-gold plasmons via charge injection in graphene. The temporal response of the metasurface was characterized, showing an ultrafast response (~200 fs), limited by the graphene plasmon lifetime and the duration of the pulses. The nonlinear response was theoretically modeled considering the near-fields on the metasurface, and the model confirms the measured enhanced tunability caused by graphene plasmons.

Previous experimental studies on graphene have focused on tunable nonlinearities in the near-infrared and visible ranges using third-harmonic generation or four-wave mixing [24, 26]. The tuning mechanism is due to changes in the one or multiphoton transition probabilities as the Fermi level is adjusted, according to the Pauli exclusion principle. Here, we leverage the nonlinear response of graphene by the enhanced near-field at the graphene plasmon resonance. However, due to damping by optical phonons, plasmons in graphene are limited to Terahertz and mid-infrared frequencies down to 6.4 µm [37]. To convert mid-infrared photons into visible light by leveraging graphene plasmons, one must employ either a four-wave mixing scheme, as shown here, or high harmonic generation. The latter is expected to be less efficient since it relies on much weaker higher-order nonlinearities. To the best of our knowledge, our work is the first experimental demonstration of coherent visible light generation using tunable graphene plasmons.

Potential applications of the metadevice and the frequency-conversion scheme introduced here include nonlinear sensing, ultrafast information processing, and ultrafast mid-infrared imaging. Since plasmon-polariton excitations in graphene can cover mid-infrared, Terahertz (surface plasmon polaritons), and microwave (spoof surface plasmon polaritons [39]) regimes, quantum transduction can potentially benefit from our approach. In principle, low-energy microwave photons could be converted into higher-energy optical photons at terahertz rates for implementation in quantum networks [40].

## Methods and Materials

*Material Stack and Sample Fabrication*

A Graphenea S11 chip was purchased, which provided the material stack as well as the microstructure. The material stack is p-doped silicon with a 90 nm layer of silica ($SiO_2$) and a monolayer of CVD grown graphene is transferred on top of the silica. Graphenea uses a photolithography process to segment the graphene into patches and to make gold contact pads on these patches. We made use of graphene patches with an area of 500 x 500 $\mu m^2$, and created patches of nanorods with an area of 250 x 250 $\mu m^2$.

To create the nanostructured gold rods, we used an electron beam lithography process. A bilayer PMMA resist was spun on with same spin parameters for both layers (4000 RPM for 60 s with 500 RPM/s ramp). The bottom layer was 2% solids 495K molecular weight PMMA in anisole solution (Kayaku PMMA 950K A2), and the top was 4% solids 950K molecular weight PMMA in anisole solution (Kayaku PMMA 950 A4). Each layer was baked at 180 C for 90 s. A 100 keV beam (Elionix ELS-G100) exposed the resist to form a deposition mask, with a beam current of 100 pA with a 1 nm shot pitch and a dose time of 0.13 $\mu$s per shot (1300 $\mu C/cm^2$). The sample was developed with a 1:3 MIBK:IPA solution for 60 s followed by an IPA bath for 30 s. E-beam evaporation was then used to deposit the metals and form the rods, consisting of 3 nm of Cr (adhesion layer, 0.3 Å/s deposition rate) followed by 25 nm of Au (0.5 Å/s). The resist and excess metals were removed by bathing the sample overnight in acetone at 45 C in a sealed beaker.

Finally, the chip was attached to a printed circuit board and the gold pads were wire bonded using an electrically conductive epoxy (EPO-TEK, H20E). The curing was carried out around 120 C in ambient conditions.

*FWM Experiment*

A Ti:Sapphire chirped pulsed amplifier generates a femtosecond pulsed 800 nm beam at a 1 kHz repetition rate. This beam is used both for the 800 nm beam in the experiment and to generate a tunable mid-infrared beam (4 to 10.5 $\mu$m) using an optical parametric amplifier followed by a noncollinear difference frequency generator. Both beams are routed to the sample and focused, and both beams strike the sample at nearly normal incidence with a small angle between the beams. The time delay of the 800 nm pulse is controlled by a retroreflector on a stage and can be adjusted until the 800 nm pulse is temporally overlapped with the mid-infrared pulse, as evidenced by the FWM signal reaching a maximum. The reflected visible FWM signal is collected by a parabolic mirror and routed to either a spectrometer or an electron-multiplying charge-coupled device for detection and analysis. Various filters are employed to control the beam power, to filter out the undesired FWM signal in the blue ($2\omega_{800nm} - \omega_{MIR}$) (long-pass filter 515 nm), and to filter out the residual 800 nm beam (short-pass filter 785 nm). A half-wave plate is used to rotate the polarization of the 800 nm beam, and a polarizer is placed before the detectors to select the detected polarization. A full diagram of the optical configuration and accounting of all components can be found in the Supplementary Information (Supp. Fig. 1).

The beam waists were measured with a knife-edge technique as 50 $\mu$m for the 800 nm beam. Measurements were also taken for the mid-infrared beam at 4000 nm (110 $\mu$m waist) and at 11000 nm (150 $\mu$m waist). The 800 nm beam should therefore limit the area of FWM signal generation, and was used to target the sample. The metasurface is 250 by 250 $\mu m^2$, and the unstructured graphene area (next to the metasurface on the same graphene patch) was 125 by 500 $\mu m^2$, so the 800 nm beam has a beam waist that fits within both samples.

*Electrical Control and Characterization*

The gate voltage of the device was applied with a source-measure unit (Keithley 2400) while measuring the source-drain current at a source-drain voltage of 1 mV (Keithley 2401). During nonlinear measurements, the gate voltage was toggled between two values (typically +40 and -20 V), and the source-drain current was noted. Separately, a transfer curve measuring the source-drain current as a function of the gate voltage was obtained. The transfer curve can be used in conjunction with the measured source-drain currents to calculate the Fermi energies as 0.14 eV (at +40 V) and 0.43 eV (at -20 V). See Supplementary Information for more details on this calculation and for the transfer curve data.

*Signal Acquisition and Processing*

The Four Wave Mixing signal was generally captured on the EMCDD camera (Andor iXonEM+ 885). The signal strength varied greatly across the spectral region of 4 to 10.5 $\mu m$ (mostly due to the power drop-off of the NDFG at higher wavelengths), and weak signals could be compensated with electron multiplying gain, lengthened exposure times, and pixel binning. Stronger signals had reduced exposure times, binning sizes, and disabled electron multiplication to avoid pixel saturation. Multiple exposures were accumulated over a duration of typically 20 to 40 seconds to average out laser fluctuations. In some cases, multiple images (2 to 8) were taken and averaged out.

When imaging the metasurface, a FWM signal that responded to a change in voltage was spatially separated from an unchanging signal, likely arising from the substrate, and the tunable signal pixels were selected for analysis. A set of nearby background pixels was also chosen, and the average value of the background pixels was subtracted from all signal pixels. The sum of the adjusted signal pixels was taken as the signal value. For graphene (or the substrate) the image was a simple spot. In this case, the strongest pixels were chosen as the signal, and the same background subtraction was performed using a set of nearby pixels. In both cases, a consistent number of signal pixels were chosen across a set of measurements.

*Measurements of the Linear Transmission Spectrum*

The linear transmission spectra of the device were found using a Fourier Transform Infrared (FTIR) spectrometer (Newport MIR8035) coupled to a mercury cadmium telluride detector (Newport 80026 MCT). The sample was illuminated by a broadband SiC mid-infrared light source (Newport 80007 SiC) and focused with a gold parabolic mirror to a spot size comparable to that of the metasurface array and the light was collected with a ZnSe focusing lens. A ZnSe wire grid polarizer selected the light polarized along the long axis of the gold rods. A spectrum of the thermal background is taken and is subtracted from all other spectra taken.

*Numerical Simulations*

Simulations were carried out using the Lumerical (Ansys) Finite Difference Time Domain (FDTD) simulation package. Graphene was modeled using a two-dimensional conductivity model [41]. The optical constants of silicon, silica thin film, and gold were taken from Palik handbooks [42]. In the mid-infrared, a scattering rate of 0.02 eV was used, which was chosen as it fit well with previous linear experiments we carried out on similar graphene samples. For the visible and near-infrared simulations, a scattering rate of 0.2 eV was chosen, as it is believed that interaction with the optical phonon of graphene will result in a high degree of damping. Simulations for the overlap integral and the linear extinction response were carried out with a broadband plane wave covering 4 to 16 $\mu$m under normal angle incidence, and the polarization was either parallel or perpendicular to the long axis of the gold rods. To calculate the temporal response as in Fig. 4(c), a narrow band pulse centered on 7.3 $\mu$m was used with a pulse duration matching that of the measured experimental pulse duration for the 8 $\mu$m laser pulse (68 fs).

Acknowledgements

Research was sponsored by the Air Force Office of Scientific Research and was accomplished under Grant Number W911NF-23-1-0156. The views and conclusions contained in this document are those of the authors and should not be interpreted as representing the official policies, either expressed or implied, of the Air Force Office of Scientific Research or the U.S. Government. The U.S. Government is authorized to reproduce and distribute reprints for Government purposes notwithstanding any copyright notation herein. Support for this project was provided by a PSC-CUNY Award, jointly funded by The Professional Staff Congress and The City University of New York. This work was partially conducted at the nanofabrication facilities of the Advanced Science Research Center at the graduate center of CUNY. The authors acknowledge Samuel Eisenberg for helping to build the experimental apparatus.